\documentclass[twocolumn]{aastex631}

\usepackage{amsmath}
\usepackage{appendix}
\usepackage{bbding} 
\usepackage{pifont}
\usepackage{bm}
\usepackage{xcolor}

\shorttitle{Multi-Epoch X-Ray study of SLSN-I 2018bsz}

\shortauthors{Ahlvind et al.}

\graphicspath{{./}{figures/}}

\begin{document}

\newcommand{\KTHOKC}{Department of Physics, KTH Royal Institute of Technology, The Oskar Klein Centre, AlbaNova, SE-106 91 Stockholm, Sweden}
\newcommand{\STHLM}{Department of Astronomy, Stockholm University,The Oskar Klein Centre, Albanova, SE-106 91 Stockholm, Sweden}

\newcommand{\lr}[1]{\left(#1\right)}
\newcommand{\vdag}{(v)^\dagger}
\newcommand\aastex{AAS\TeX}
\newcommand\latex{La\TeX}
\newcommand{\jj}[1]{\textcolor{red}{X: #1}}

\title{Multi-Epoch X-Ray Detection of SLSN-I 2018bsz: Constraints on the Powering Mechanism and Ejecta Structure} 

\correspondingauthor{Julia Ahlvind}
\email{ahlvind@kth.se}

\author[0009-0002-2740-9570]{Julia Ahlvind}
\affiliation{\KTHOKC}

\author[0000-0003-0065-2933]{Josefin Larsson}
\affiliation{\KTHOKC}

\author[0000-0002-0427-5592]{Dennis Alp}
\affiliation{\KTHOKC}

\author[0000-0001-9454-4639]{Ragnhild Lunnan}
\affiliation{\STHLM}

\received{April 09, 2026} \revised{ May 29, 2026} \accepted{June 11, 2026}
\submitjournal{ApJ}

\begin{abstract}
SN~2018bsz is the closest known stripped superluminous supernova (SLSN-I) to date, making it an ideal laboratory for investigating the physical mechanisms powering this class of extreme explosions. We present a multi-epoch X-ray spectroscopic study of SN~2018bsz based on four Chandra observations followed by one XMM observation, spanning 87 to 1253~days after explosion. The source is detected at all Chandra epochs and is also tentatively detected in the late XMM observation, although more uncertain due to nearby contaminating sources. Regardless of the XMM detection, this makes SN~2018bsz the second X-ray detected SLSN-I and the third X-ray detected SLSN overall. We explore potential power sources for the observed X-ray emission and find that a millisecond magnetar central engine underpredicts most of the observed X-ray luminosities and fails to reproduce the relatively flat light curve. Accounting for ejecta absorption further increases the discrepancy. While asymmetries and magnetar-driven ionization could reduce the effective absorption, ionization breakout is expected years after our observational window. Instead, the observations are more readily explained by early-time interaction between the ejecta and the circumstellar medium, while the magnetar emission is absorbed by the ejecta. This scenario is supported by the flat temporal evolution, previous optical results, and inferred mass-loss rates which resemble those of stripped supernovae that later evolve into interacting systems. Our results thus favor the scenario where SN~2018bsz is part of a distinct group of SLSN-I, where interaction is crucial for the strong emission.

\end{abstract}


\section{Introduction} \label{sec:intro}
\setcounter{footnote}{2}

Superluminous supernovae (SLSNe) are rare types of stellar explosions with energies $\sim100$ times that of ordinary supernovae (SNe). The mechanism producing the excess energy compared to regular core-collapse SNe remains debated. Commonly proposed explanations include strong interaction between the circumstellar medium (CSM) and ejecta \citep{2008Smith,2017Inserra}, highly magnetized and rapidly rotating neutron stars (magnetars; \citealt{2007Maeda,2010Kasen,2010Woosley}), accreting black holes \citep{2013Dexter}, or a combination of different mechanisms. The former mechanism is typically invoked for hydrogen rich SLSNe (SLSNe-II), while the latter two are often proposed for hydrogen-poor SLSNe (SLSNe-I). 

Using a magnetar as the driving force of SLSN-I, the observed spectra and light curves have been successfully modeled \citep{2012Dessart,2013Howell,2016Mazzali,2017Nicholl,2024Gomez}. However, some SLSNe-I show emerging hydrogen emission lines at later times \citep{2015Moriya,2015Yan,2016Nicholl,2017Yan}, suggesting interaction between ejecta and previously expelled hydrogen layers. This transition has also been observed in ordinary stripped SNe \citep{2006Soderberg,2015Milisavljevic,2018Chen,2018Kuncarayakti,2018Mauerhan,2020Sollerman}. The strong shift from stripped to interacting SNe suggests a shell-like CSM structure that could come from eruptive mass loss of the progenitor star, rather than from a steady wind. These bursts of mass loss is common in luminous blue variable (LBV) stars, which are known to be the progenitors of initially strongly interacting SNe (Type IIn) \citep{2014Smith}, and may also explain the superluminous population \citep{2011Smith,2014Justham}. Furthermore, \cite{2025Gkini} recently showed that such shell-like CSM structures are seen for a sample of SLSNe-I and is consistent with mass-loss episodes from LBV-like eruptions or pulsational pair-instability events.

Despite the successful modeling of optical data, distinguishing between the power sources remains challenging as different mechanisms can produce similar optical signatures (e.g., \citealt{2018DeCia,2023Chen}). Observations at other wavelengths are therefore essential to break this degeneracy and probe the physical conditions of interaction or a central engine more directly. Some studies have addressed this by using radio observations and detected SLSNe with support of both types of power origin \citep{2019Eftekhari,2023Margutti}.

X-ray observations provide another important probe of both CSM interaction and central engine activity.
Detecting X-rays associated with magnetar activity requires careful timing. At early epochs, the dense ejecta efficiently absorb soft X-rays, but as the ejecta expand over time, the absorption and the optical depth ($\tau$) decrease \citep{2018Alp,2025Ahlvind}. Observations must therefore be sufficiently late for the ejecta to have become partially optically thin, yet early enough for the decaying magnetar luminosity to remain detectable. In addition, magnetars, with their associated pulsar wind nebulae (PWNe), may be able to ionize the ejecta, thereby making the medium transparent to X-rays \citep{2014Metzger}. This could make the magnetar luminosity detectable at earlier epochs. Late-time detections or limits of X-ray emission, some months to years after explosion, would offer valuable constraints on magnetar properties.

While some work on X-ray emission from SLSNe has been presented \citep{2007Smith,2013Levan,2013Ofek,2018Margutti}, the detections often cover early epochs reaching only $\sim160$~days post explosion. The majority of the X-ray observations of SLSNe result in nondetections. In most cases, the observations were obtained close to the optical peak and/or are too shallow to probe the X-ray luminosities expected from CSM-shock interaction or emerging magnetar emission (e.g., \citealt{2013Ofek}). While upper X-ray luminosity limits may constrain the mass loss rates \citep{2013Levan}, late detections are required to identify the powering mechanism. To date, only one X-ray detection of an SLSN-I has been claimed: the SLSN-I SCP~06F6 at $\sim150~$days after explosion with an X-ray luminosity of $\sim1\times10^{45}~\rm erg~s^{-1}$ in the 0.2--10~keV energy band \citep{2013Levan}. \cite{2018Margutti} also report an X-ray detection which could be associated with the SLSN PTF~12dam, but state that the star formation within the host galaxy partly contributes to the X-ray luminosity and treat the result as an upper limit. Similarly, only one SLSN-II has been detected in X-rays; SN~2006gy measured to a luminosity of (0.5--2~keV) $1.65\times10^{39}\rm~erg~s^{-1}$ at $\sim 60~$days post explosion \citep{2007Smith}. This SLSN persistently showed narrow lines in its spectrum, and was originally classified as a Type IIn \citep{2007Smith,2007Ofek}. The narrow lines and exceptional luminosity suggest that SN~2006gy could be considered an SLSN-IIn (e.g. as mentioned in \citealt{2014Quimby,2020Nyholm}).

In this study, we present an in depth X-ray analysis of SN~2018bsz, theclosest observed SLSN-I to date at 111~Mpc. SN~2018bsz was discovered on May 17th 2018 \citep{2018Stanek,2018Brimacombe} by the All Sky Automated Survey for Supernovae\footnote{\url{http://www.astronomy.ohio-state.edu/~assassin/index.shtml}} (ASAS-SN; \citealt{2014Shappee}). The explosion date is estimated to be 2018 March 25 by \cite{2018Anderson}, who also reclassified it from a Type II SN to an SLSN-I \citep{2018aAnderson}. Although SN~2018bsz shows common features with other SLSNe-I, such as strong C~II and O~II absorption features in the early spectra \citep{2018Anderson} and a late-time spectral evolution resembling that of ordinary stripped Type Ic, it also shows a number of notable deviations. Firstly, the optical light curve exhibits a long and slow rise, with a sudden steepening just before peak brightness, followed by hydrogen emission lines emerging $\sim25~$days after peak \citep{2022Pursiainen}. SN~2018bsz is also the first SLSN to show dust formation within the ejecta \citep{2021Chen}. Although hydrogen lines have previously been detected for a small sample of SLSNe-I \citep{2015Moriya,2015Yan,2016Nicholl,2017Yan}, SN~2018bsz is unique in exhibiting a multi-component H$\alpha$ profile, a characteristic more commonly associated with SN ~IIn \citep{2022Pursiainen}. Based on this line profile and its spectral evolution, \cite{2022Pursiainen} proposed that this event is interacting with an asymmetric, disk-like CSM.

In this paper, we study XMM and Chandra observations of SN~2018bsz from $\sim3$~months to $\sim3$~yr post explosion. 
The proximity and multi-epoch X-ray observations of SN~2018bsz make it a well-suited target for investigating the high-energy emission from SLSN-I and for placing constraints on proposed powering mechanisms, including magnetar central engines and CSM interaction. 
The paper is organized as follows. Section~\ref{sec:data_reduction} presents the observation and data reduction. The analysis and main results are described in Section~\ref{sec:results_and_analysis}. In Section~\ref{sec:discussion}, we discuss the results and investigate the origin of the X-ray emission. We finally present the conclusions and summary in Section~\ref{sec:summary_conclusion}.

\section{Observations and Data Reduction} \label{sec:data_reduction}
This study makes use of all available X-ray observations of SN~2018bsz. This includes four Chandra observations near the time of explosion (87--304~days post explosion, PI Margutti) and one later XMM-Newton observation (1253~days post explosion, PI: Alp) (see Table~\ref{tab:observaions}).

\begin{deluxetable*}{l l l l l l l l l}
\tablenum{1}
\tablecaption{X-Ray Observations \label{tab:observaions}}
\tablewidth{0pt}
\tablehead{ 
 \colhead{ObsID}& \colhead{Telescope} &\colhead{Obs. Date}  & \colhead{Epoch} & \colhead{$t_{\rm exp}$}  & \colhead{Sig.} & \colhead{Net Counts} & \colhead{src (RA, DEC)} & \colhead{Source Region Radius}\\
 \colhead{}  &  \colhead{}  & \colhead{}& \colhead{(days)} & \colhead{(ks)} &\colhead{} & \colhead{} & \colhead{} & \colhead{(arcsec)}
}
\startdata
        20289 & Chandra & 2018-06-20 & 87 & 49.62 & 3.2 & 10 & 16:09:39.0816, -32:03:46.394 & 2.530\\
        20290 & Chandra & 2018-08-01 & 130 & 25.92 & 6.9 & 17& 16:09:39.0719, -32:03:46.301 & 1.820\\
        21665 & Chandra & 2018-08-14 & 143 & 20.75 & 3.1 & 9 & 16:09:39.0912, -32:03:45.643 & 2.688\\
        20291 & Chandra & 2019-01-22 & 304 & 49.40 & 4.4 & 12 & 16:09:39.0725, -32:03:45.535 & 2.000\\
        0882400101 & XMM EPIC-PN & 2021-08-29 & 1253 & 15.46 & 4.4 & 26 & 16:09:39.1601, -32:03:48.213 & 10.000\\
        0882400101 & XMM EPIC-MOS1 & 2021-08-29 & 1253 & 25.83 & 4.0 & 15 & 16:09:39.0725, -32:03:45.535 & 12.811\\
        0882400101 & XMM EPIC-MOS2 & 2021-08-29 & 1253 & 24.68 & 3.9 & 8 & 16:09:38.9116, -32:03:45.762 & 10.000\\[2mm]
\enddata
\tablecomments{\footnotesize{$\rm t_{exp}$ is the effective time after background flare correction (only relevant for XMM, see Section~\ref{sec:data_reduction_xmm}) and Sig. is the image detection significance. The XMM source region includes weak sources located close to SN~2018bsz, discussed further in Section~\ref{sec:results_and_analysis}. The coordinates were obtained from \texttt{wavdetect} for Chandra and \texttt{ewavelet} for XMM.
}}
\end{deluxetable*}

\begin{figure*}
\gridline{\fig{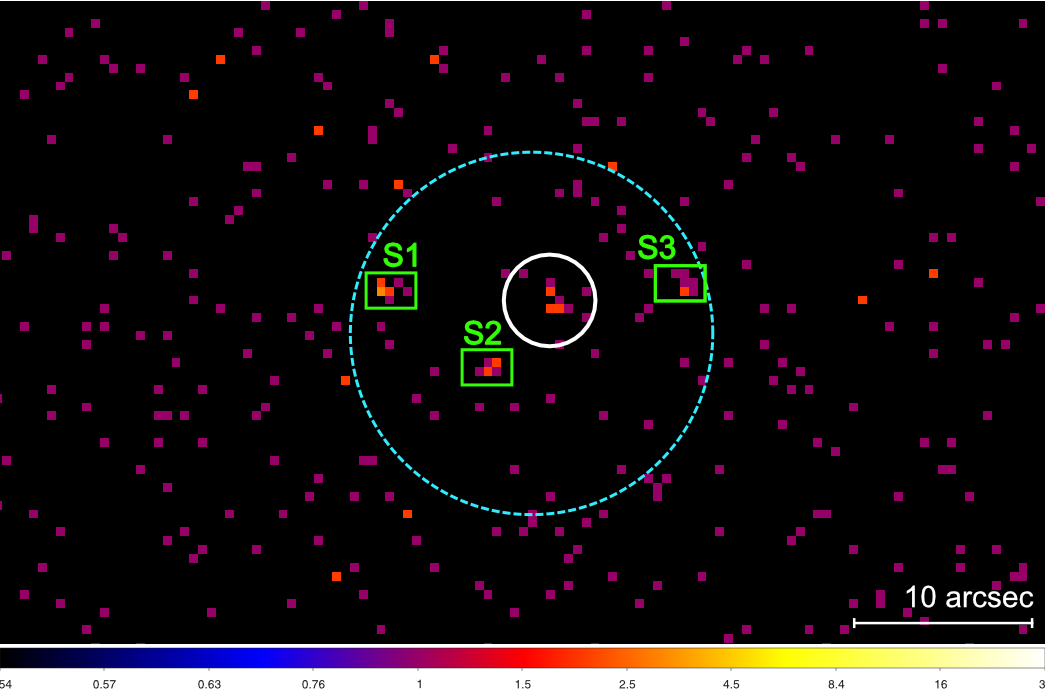}{0.5\textwidth}{(a) Chandra ObsID=20289, epoch=87~days}
          \fig{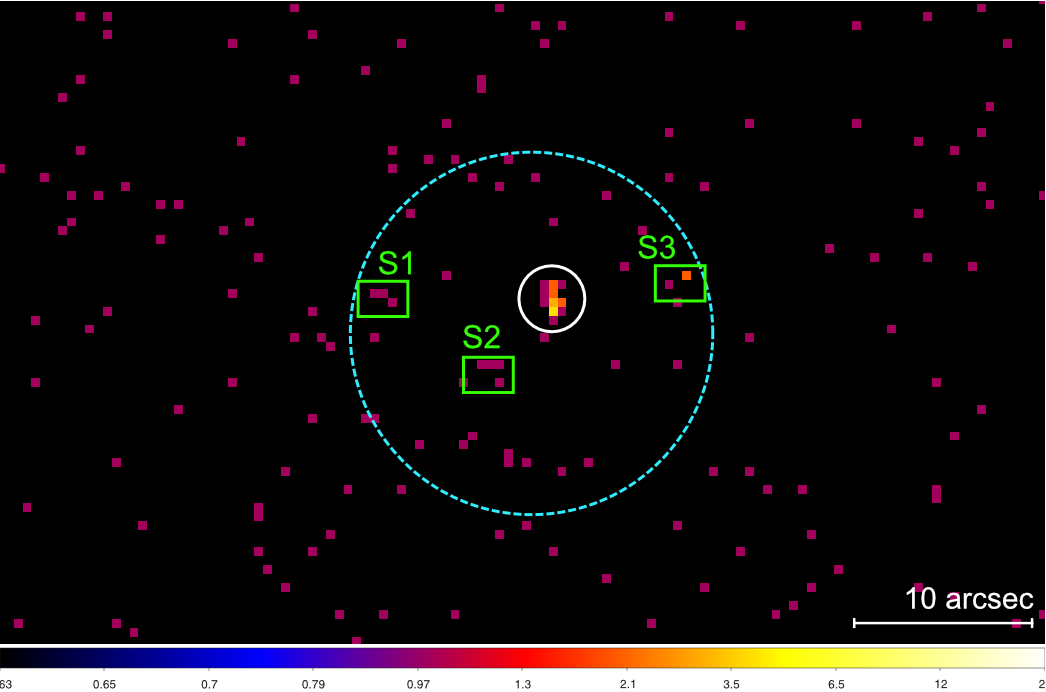}{0.5\textwidth}{(b) Chandra obsid=20290, epoch=130~days}
          }
\gridline{\fig{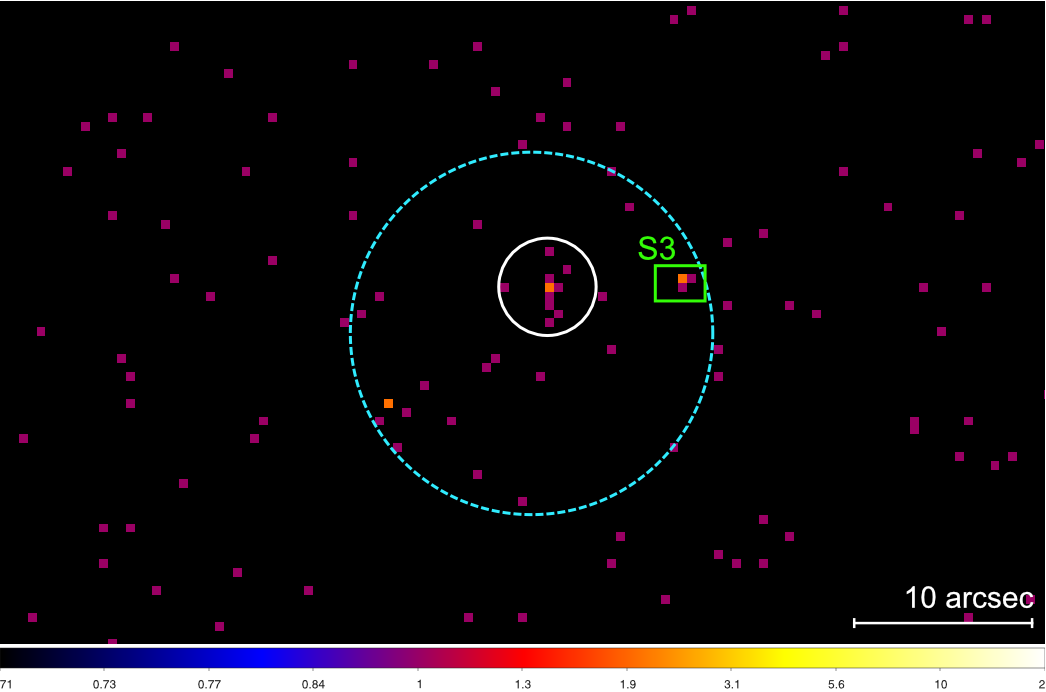}{0.5\textwidth}{(c) Chandra ObsID=21665, epoch=143~days}
          \fig{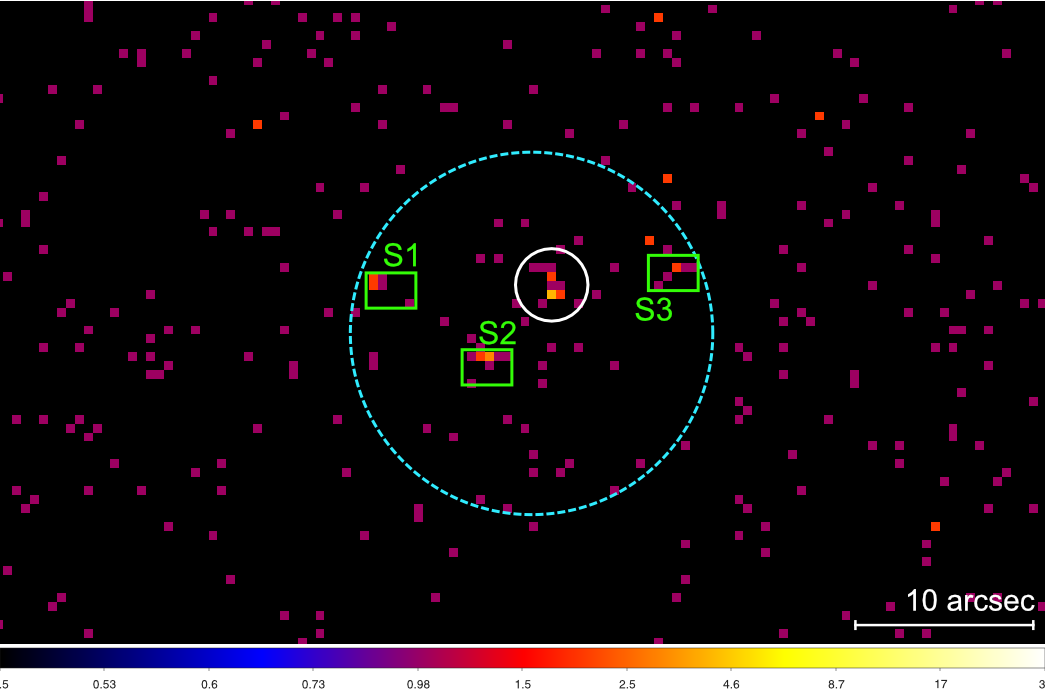}{0.5\textwidth}{(d) Chandra obsid=20291, epoch=304~days}
          }
\gridline{\fig{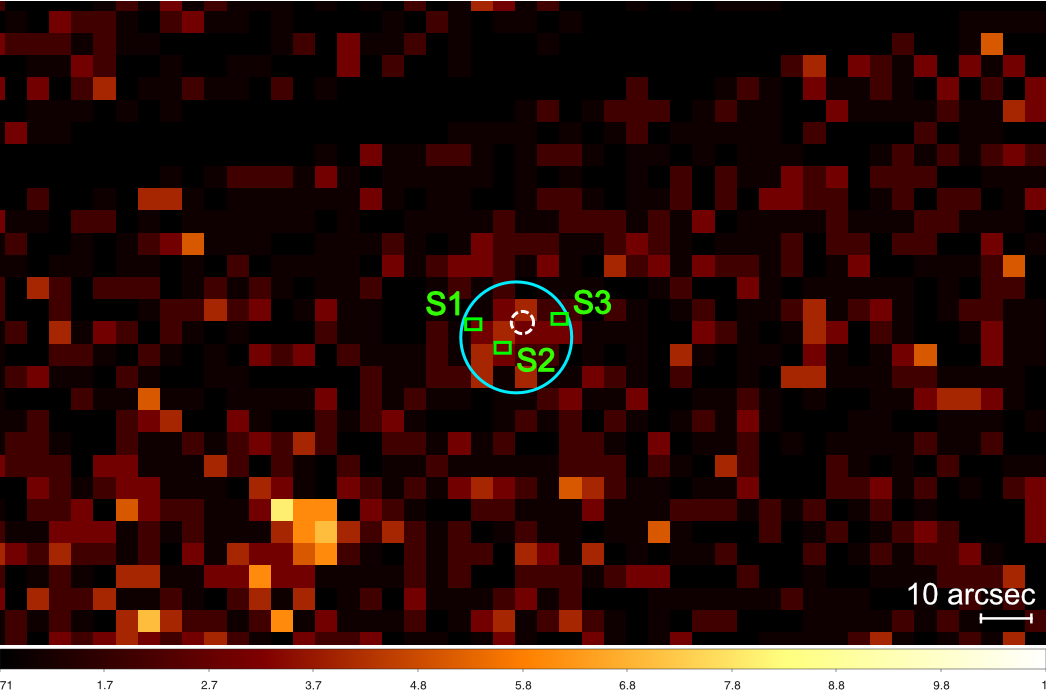}{0.5\textwidth}{(e) XMM ObsID=0882400101, epoch=1253~days}
        \fig{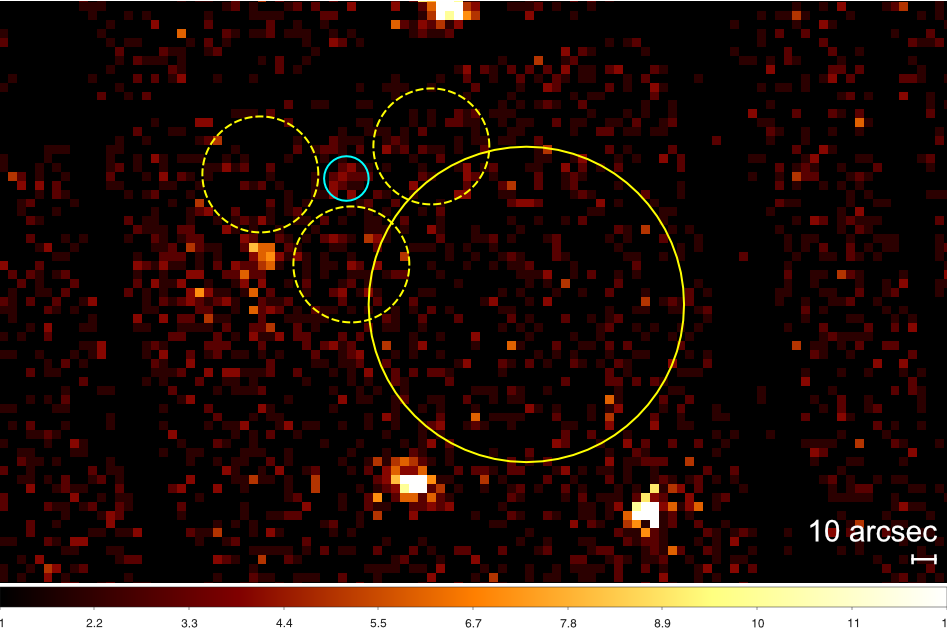}{0.5\textwidth}{(f) XMM ObsID=0882400101, epoch=1253~days}}
\caption{(a)--(d): Chandra images in the 0.5--8~keV energy range. (e) and (f): XMM EPIC-pn image in the 0.5--10~keV energy range. In panels (a) to (e) the solid circles indicate the source regions used in each observation, while dashed circles mark the reference source regions from the other telescope. In the Chandra panels (a--d), the solid white regions correspond to Chandra source and the dashed cyan circles to the equivalent XMM EPIC-pn source region; in the XMM panel (e), the solid cyan region corresponds to XMM EPIC-pn source region and the dashed white circle indicates the Chandra source region from observation 20291. The additional X-ray sources (S1--S3) close to the SN position in the Chandra images are also marked by green rectangles in panels (a)--(e). The color bar in each panel corresponds to the count rate in each pixel. Note the different field of views (FoV) for the Chandra and XMM images, as indicated by the scale bars in the lower right corners. Panel (f) shows the XMM image over a larger FoV, with corresponding source region (cyan) and background region (solid yellow) used for spectral extraction. The dashed yellow circles are alternative background regions that were investigated.
\label{fig:obs_images}}
\end{figure*}

\subsection{Chandra}\label{sec:data_reduction_chandra}
SN~2018bsz was observed in four Chandra observations, each carried out with the ACIS-S instrument, specifically using chip S3. The data were downloaded from the Chandra archive and processed with CIAO ver. 4.13 \citep{2006Fruscione}. Event files and images were produced through the CIAO task \texttt{chandra\_repro} and \texttt{dmcopy}.

Images (bin=1) in the 0.5--8~keV energy range were used for running a source detection algorithm. Point-source identification employed the wavelet-based algorithm \texttt{wavdetect}, which correlates the image with wavelets of different scales (radii 1.0, 2.0, and 3.0 pixels). A significance threshold of \texttt{sigthresh=$10^{-5}$} was adopted, and standard parameters were retained otherwise. A detection of $>3\sigma$ at the positions consistent with the optical coordinates of the SNe is seen in all Chandra images (see Table~\ref{tab:observaions} for more details) . We also detect additional sources within 10$\arcsec$ from the SN position (see Figure~\ref{fig:obs_images}). These additional sources could potentially contaminate the SN spectrum of the XMM observation, and are therefore analyzed as described later in Section~\ref{sec:results_and_analysis}.  

Source and background regions were defined manually after visual examination of the 0.5--8~keV images. The circular source regions typically have radii of $\sim3\arcsec$ (see Table~\ref{tab:observaions} for details), while background regions of $\sim~23\arcsec$ were chosen close to the source but clear of unrelated emission. Manual definition of the source regions and sizes was preferred, as the automatic CIAO task \texttt{psfsize\_srcs} often returned unrealistically small regions that excluded a great portion of the SN emission.

Light curves were created using \texttt{dmextract} to inspect and filter out intervals affected by background flares using \texttt{deflare}. As the light curves showed no significant flaring events, this had little to no impact on the observation and effective exposure time.

Spectral products of each observation were extracted using \texttt{specextract}. The auxiliary response files (arf) were generated without weighting, which is appropriate for point-source analysis, but with a point-source aperture correction applied to the unweighted arf. Owing to the limited number of counts, Poisson-based C-statistics was used for spectral analysis in XSPEC. To accommodate this, Chandra spectra were grouped with one count per spectral bin using \texttt{grpmincounts=1} within \texttt{spacextract}. 

\subsection{XMM}\label{sec:data_reduction_xmm}
The XMM EPIC datasets were retrieved from the XMM Science Archive and processed using the calibration files (CCF) from 2025. Data reduction followed the standard guidelines provided for the Science Analysis Software (SAS, v22.1.0; \citealt{2014heasoft}). Images with a bin size of one were produced over the 0.5-–10~keV energy range and used to perform source detection.

Source detection was carried out with the SAS task \texttt{ewavelet}, which identifies significant wavelet correlations across multiple spatial scales. This approach closely parallels the \texttt{wavdetect} algorithm employed for Chandra. We find a significant detection centered $3.6\arcsec$ from the optical SN position. This offset is discussed in more detail in Section~\ref{sec:results_xmm}, but likely comes from the additional aforementioned X-ray sources seen in the Chandra images. Nevertheless, the optimal source region, described below, is large enough to enclose the whole source (see Figure~\ref{fig:obs_images}).

Source regions for the XMM observations were determined using the SAS task \texttt{eregionanalyse}, which calculates the circular radius (between $\sim10-13\arcsec$) and maximizes the signal-to-noise ratio, accounting for the PSF fraction, source counts, and local background. Background regions were selected by visual inspection (see solid yellow circle in panel (f) of Figure~\ref{fig:obs_images}), placed as close to the target as possible and always on the same CCD, while avoiding contaminating sources. They typically have radii between five and eight times larger than the associated source region. We confirm that there is negligible contamination of the source region from diffuse gas in the host galaxy by trying out different background regions (dashed yellow circles in panel (f) of Figure~\ref{fig:obs_images}) that are smaller, but closer to the source region. The number of counts per arcsec$^2$ in all four background regions are similar and the net count rate for the source region ranges from 24--27 in EPIC-PN, with the different backgrounds.

Source and background spectra were extracted from filtered event files in which high-background intervals had been excluded. The filtering was based on the 10-–12~keV light curves, removing time periods where the count rate exceeded 4 counts$~\rm s^{-1}$ for EPIC-PN and 3.5 for EPIC-MOS, where the EPIC-PN observation is the most affected. The total exposure time for the XMM observation is 26.6~ks, while the remaining good exposure times are $\sim 15$~ks for PN and $\sim 25$~ks for MOS (see Table~\ref{tab:observaions} for details). Source spectra were extracted with aperture corrections and redistribution matrix (rmf) weighting applied using the default SAS settings optimized for point sources.

Spectral binning was performed with \texttt{grppha}, adopting one count per bin for all EPIC spectra (PN, MOS1, and MOS2), utilizing C-statistics for low count spectra. 

\section{Analysis and Results}\label{sec:results_and_analysis}

In this study, we investigate the nature of the powering source of SN~2018bsz. The spectra are therefore fitted with both a power law, suggesting a millisecond (ms) pulsar origin, and a plasma model, representing an interaction driven power source. All fits include Galactic absorption that is modeled by the XSPEC component \texttt{tbabs} \citep{2000Wilms}, where the abundance is set according to \cite{2000Wilms} and the column density of hydrogen is fixed at $\rm NH=2.02\times 10^{21}~cm^{-2}$, obtained from the NHtot tool\footnote{\url{https://www.Swift.ac.uk/analysis/nhtot/}} \citep{2013Willingale}. Additional fits including intrinsic absorption from the host galaxy resulted in column densities consistent with zero in all model configurations described in the following sections and were therefore excluded from the final fits. 

All spectral analysis is performed using XSPEC ver. 12.15.0c \citep{1996Arnaud} with the C-stat fit statistic. Chandra spectra are fitted within 0.5--8~keV, and similarly the fluxes and luminosities are derived within this interval. For the luminosities used in the pulsar analysis (Section~\ref{sec:pulsar_discussion}) we extrapolate the model fits up to 10~keV and derive luminosities in the 2--10~keV energy range. XMM spectra are fitted within 0.5--10~keV and luminosities are derived in the same interval. All fluxes and luminosities presented in this work have been corrected for Galactic absorption and are presented with 90\% error bars, or as $3\sigma$ upper limits in Tables~\ref{tab:results_chandra} and \ref{tab:results_xmm}. Below we describe the analysis methodology and present the results for Chandra and XMM in sections \ref{sec:results_chandra} and \ref{sec:results_xmm}, respectively. 

\subsection{Chandra}\label{sec:results_chandra}

A clear X-ray source is detected ($> 3\sigma$) in all four Chandra observations, located typically $\sim0.3\arcsec$ from the updated SN coordinates \citep{2018Anderson} (up to $\lesssim1\arcsec$ in ObsID 20289).  The 1$\sigma$ positional uncertainties from \texttt{wavdetect} (the statistical centroid variances) range between 0.11--0.23$\arcsec$ for RA and 0.15--0.33$\arcsec$ for DEC for the four Chandra observations. Furthermore, the systematic astrometric error for the Chandra CSC 2.0 catalog is 0.71$\arcsec$ at 95\% confidence \citep{2024Evans}. The source is significantly detected ($>3\sigma$, see Table~\ref{tab:observaions} for details) in the 0.5--8~keV images across 87--304~days after explosion, and we associate it with SN~2018bsz. The spectra are well described by power-law models modified by Galactic absorption, with hard spectral indices ($\Gamma$) in the range 0.9--1.9 (see Table~\ref{tab:results_chandra}). Attempts to fit thermal plasma models (\texttt{mekal}) resulted in temperatures driven to the upper bound of the model grid  (78~keV), effectively reproducing a power-law shape. This behavior is expected given the scarcity of counts below 1~keV (see the binned spectra in Figure~\ref{fig:spectra} (a)).

\begin{figure*}
\hspace{0cm}\includegraphics[width=1\textwidth]{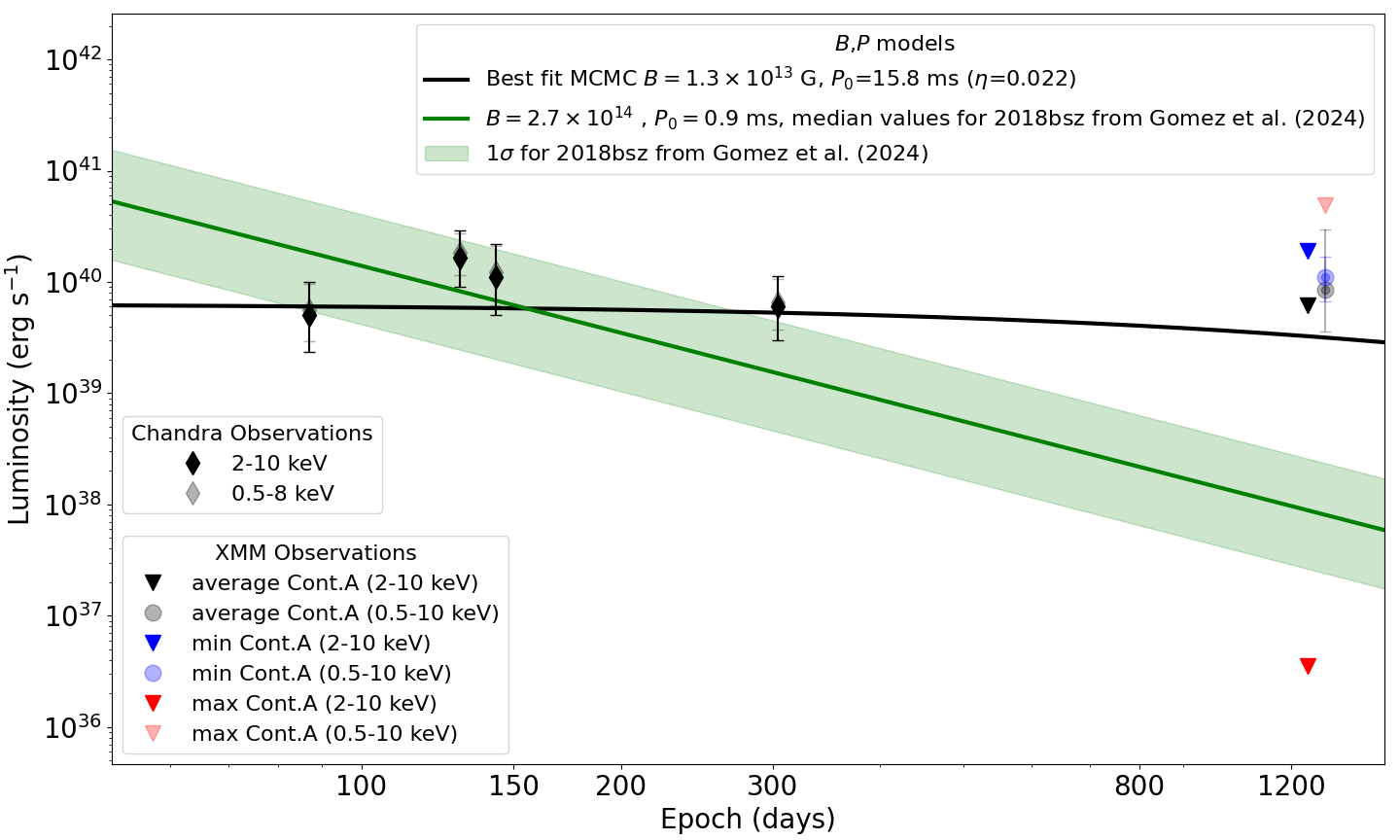} 
\caption{Magnetar evolution models compared with observational data. The black solid line shows the best-fitted $B$ and $P_0$ with fixed $\eta$ from the MCMC fit. The green line and shaded region show the pulsar model with best-fit values of $B$ and $P_0$ for the optical light curve fit of SN~2018bsz from \cite{2024Gomez}. The diamonds are the power-law luminosities from Chandra and the triangles (3$\sigma$ upper limits) and dots (detections) are from XMM. All error bars represent 90\% uncertainty. Three different colors (black, blue and red) are shown for the XMM results and correspond to the average, min Cont.A and max Cont.A, respectively. The former limit was used in the magnetar fits. Finally, the solid colors are X-ray luminosities in the energy range 2--10~keV and the more transparent 0.5--8~keV for Chandra, and 0.5--10~keV for XMM. The 0.5--10~keV XMM data points have been shifted +60~days for visual clarity. 
\label{fig:pulsar_evolution}}
\end{figure*}

Each individual observation contains a limited number of counts, and the final fits yield hard spectral indices with large associated uncertainties, which further show no significant temporal evolution (Table~\ref{tab:results_chandra}, upper panel). This motivated a simultaneous fit of the four Chandra observations, tying  $\Gamma$ across epochs while allowing the normalization to vary. The resulting model is shown in Figure~\ref{fig:spectra} (a). The corresponding luminosities, presented in Table~\ref{tab:results_chandra} (lower panel) and Figure~\ref{fig:pulsar_evolution}, are consistent with being constant over time.

\begin{deluxetable*}{l l l l l  l l}
\tablenum{2}
\tablecaption{Spectral Results for Chandra Data\label{tab:results_chandra}}
\tablewidth{0pt}
\tablehead{ 
 \colhead{ObsID}  &  \colhead{Epoch}  & \colhead{$\Gamma$} & \colhead{C-stat}& \colhead{dof}& \colhead{$L_{\rm0.5-8~keV}$} & \colhead{$L_{\rm2-10~keV}$}\\
 \colhead{}  & \colhead{(days)} & \colhead{}& \colhead{}& \colhead{}& \colhead{($10^{40}~\rm erg~s^{-1}$)} & \colhead{($10^{40}~\rm erg~s^{-1}$)}
}
\startdata
        20289 &  87  & $1.9_{-1.1}^{+1.2}$ & 24.97 & 22 & $0.49_{-0.24}^{+0.43}$ & $0.29_{-0.21}^{+0.54}$\\
        20290  & 130  & $1.26_{-0.79}^{+0.83}$ & 23.76 & 20 & $1.85_{-0.71}^{+1.10}$ & $1.8_{-1.0}^{+1.9}$\\
        21665  & 143 & $0.9_{-1.3}^{+1.3}$ & 13.33 & 8 &  $1.41_{-0.76}^{+1.46}$ &  $1.6_{-1.1}^{+2.9}$\\
        20291  & 304  & $1.4_{-1.1}^{+1.2}$ & 13.59 & 11 & $0.65_{-0.30}^{+0.51}$ & $0.58_{-0.39}^{+0.91}$\\[2mm]
        \hline
        20289 &  87 & $1.32_{-0.51}^{+0.53}$ & 48.64 & 47 & $0.56_{-0.27}^{+0.40}$ & $0.51_{-0.27}^{+0.50}$\\
        20290  & 130  &  &  &  & $1.82_{-0.67}^{+0.92}$ & $1.66_{-0.76}^{+1.24}$\\
        21665  & 143  &  &  &  &  $1.22_{-0.61}^{+0.89}$ &  $1.11_{-0.61}^{+1.08}$\\
        20291  & 304  &  &  &  & $0.66_{-0.29}^{+0.41}$ & $0.60_{-0.30}^{+0.52}$\\[2mm]
\enddata 
\tablecomments{\footnotesize{ The spectra were fitted individually in the upper panel and with $\Gamma$ tied in the lower panel. All luminosities have been corrected for Galactic absorption. Uncertainties are 90\%. The distance adopted in the luminosity calculation is 111~Mpc.
}
}
\end{deluxetable*}

Inspection of the Chandra images and \texttt{wavdetect} results revealed three fainter X-ray sources near the SN position and within the XMM SN extraction region (see green boxes S1-S3 within the cyan dashed circular regions in Figure~\ref{fig:obs_images} (a)--(d)). These sources are detected with lower- or equivalent significance as the SN in all Chandra epochs, and with fewer net counts: S1:$\sim$3--6, S2:$\sim$4--6 and S3:$\sim$2--9 compare to the SN in Table~\ref{tab:observaions}. Due to the lower angular resolution in XMM the emission from these sources will blend with the SN emission in the XMM spectra. We therefore analyze the combined S1--S3 Chandra spectra so that their contribution can be accounted for in the spectral model for the XMM fit. 

Spectral extraction for these contaminating sources is performed using an annulus defined as
the Chandra source region subtracted from the XMM EPIC-PN circular SN source region (the contamination annulus; Cont.A, illustrated by the region between the cyan dashed and white solid circles in Figure~\ref{fig:obs_images} (a)--(d)). These spectra are produced using the same procedures and background regions as for the SN spectra described in Section~\ref{sec:data_reduction_chandra}. The Cont.A spectra from all four epochs are fitted simultaneously with model parameters tied between observation, after confirming that they show no significant temporal evolution. 

We find that the spectrum of the contamination region is well described by a power-law model with best fit parameters $\Gamma_{\rm cont. A}=1.07_{-0.53}^{+0.58}$ and normalization $N_{\rm Cont. A}=8.77_{-3.97}^{+5.73}\times10^{-7}$. The true nature of S1--S3 is difficult to assess as there are limited number of multi-wavelength observations in this field. It is possible that the emission comes from the hot gas in the host galaxy, as Pan-STARRS pre-explosion images show that the system is an ongoing merger of two dwarf galaxies \citep{2018Anderson}. A caveat is that such emission is typically well described by a low-temperature thermal model \citep{2005Grimes,2009Owen,2011Boroson}, while S1--S3 is best fit with a hard power law. Alternatively, the emission could come from high mass X-ray binaries or similar.

\subsection{XMM}\label{sec:results_xmm}
Running \texttt{ewavelet} on the XMM images, we detect (at $\ge3\sigma$) a source at $\sim3.6\arcsec$ from the updated SN position \citep{2018Anderson}. The absolute astrometric uncertainty of XMM EPIC is $1.2\arcsec(1\sigma)$\footnote{\url{https://xmmweb.esac.esa.int/docs/documents/CAL-TN-0018.pdf}}, making the optical SN position and the detected X-ray source consistent within $3\sigma$. As noted above, additional X-ray sources were identified near the SN position in the Chandra images. Due to the lower angular resolution of XMM, these different sources are unresolved and identified as a single source by \texttt{ewavelet}, explaining the slightly larger offset from the SN coordinate. In particular, the algorithm places the center of the source centroid between the SN coordinate and the closest contaminating source, S2 (Figure~\ref{fig:obs_images}e). S2 is one of the brightest nearby X-ray sources and is likely to significantly contaminate the XMM spectrum, especially if the SN emission has faded since the Chandra observations. The best fit Cont.A model described at the end of Section~\ref{sec:results_chandra}, predicts 15 net counts in the XMM EPIC-PN spectrum, compared to the total number of 26 net counts in the spectrum for the energy range 0.5--10~keV.

To isolate the SN contribution and account for blended emission from the unresolved nearby sources, we introduce an additional power-law component as described at the end of Section~\ref{sec:results_chandra}. This component is included in the XMM SN fit with $\Gamma_{\rm Cont. A}$ and $N_{\rm Cont.A}$ frozen at previously determined values. Three scenarios are considered using the best fit $N_{\rm Cont. A}$ and its lower and upper 90\% error bounds, corresponding to the average, minimal, and maximal levels of contamination from S1--S3. The resulting model for the XMM fits is thus \texttt{tbabs(pow$_{\rm Cont.A}$ + pow)}, with the unabsorbed flux extracted from the second power-law component representing the SN, having free $\Gamma$ and normalization.

\begin{figure*}[ht!]
\gridline{
    \fig{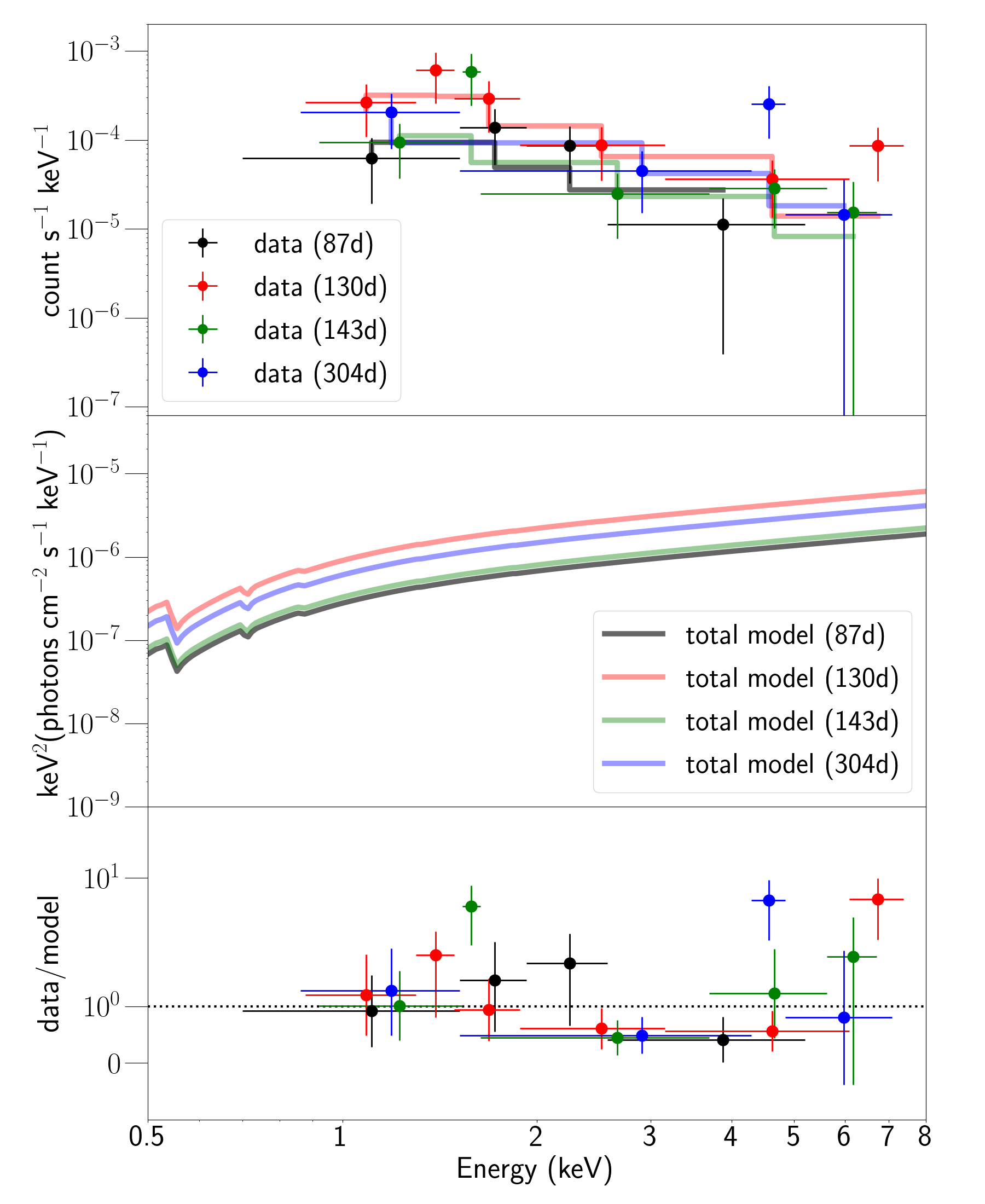}{0.5\textwidth}{(a) All Chandra spectra fitted with $\Gamma$ tied but normalization free to vary between observations. Each color represents a spectrum and corresponding model, where the epoch of each observation is noted in the label. The models shown in the middle panel consist of Galactic absorbed power laws. The spectra were binned for visual clarity to three counts per bin. }
    \fig{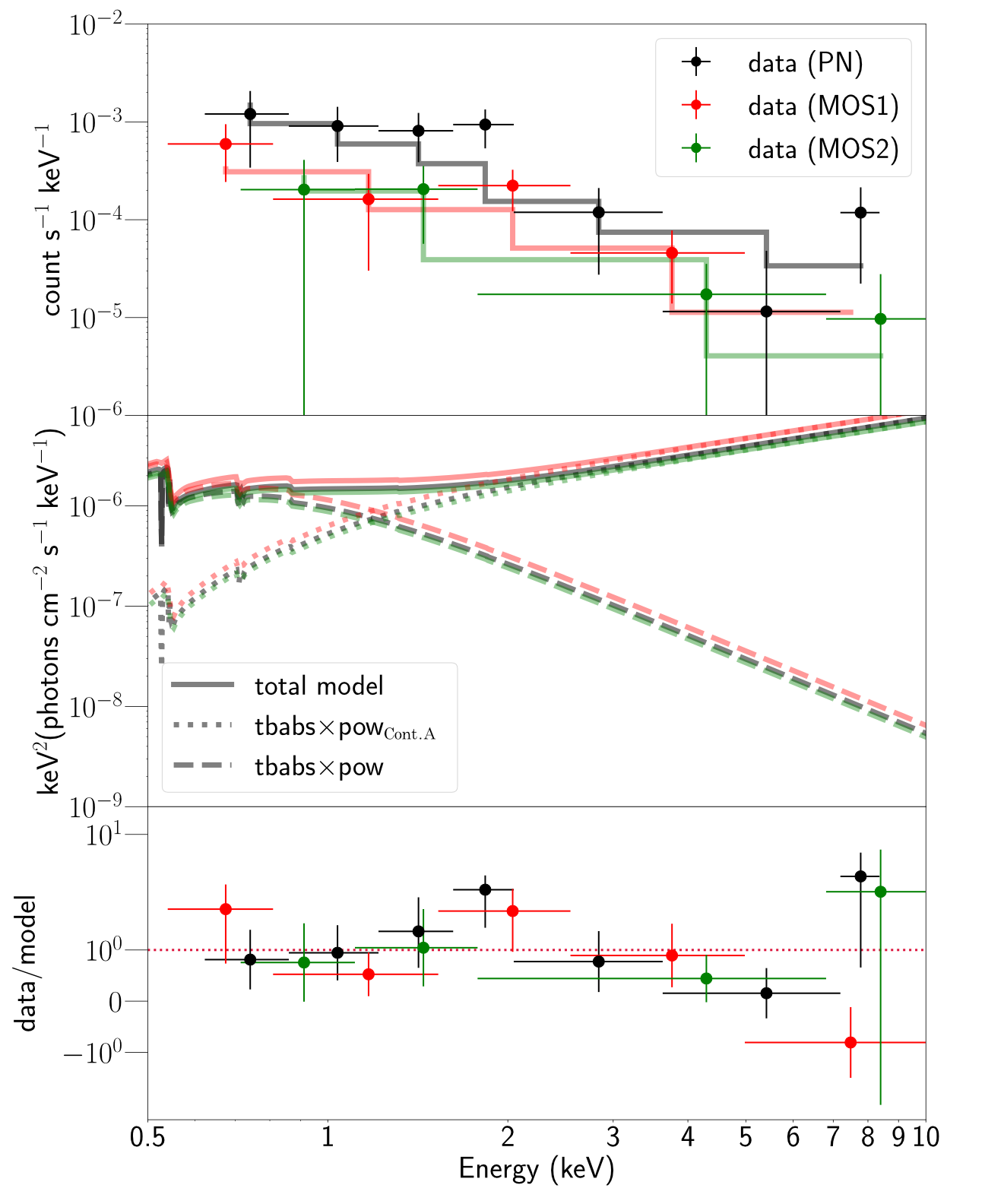}{0.5\textwidth}{(b) EPIC XMM spectra, fitted with tied $\Gamma$ and normalization, but with PN-MOS variation set to $\leq8\%$. The colors correspond to different EPIC instruments as noted in the labels of the top panel. The total models shown as solid lines in the middle panel consist of Galactic absorption, power law for Cont.A and the SN power law. The dotted lines show the Galactic absorbed Cont.A power law, and the dashed lines the Galactic absorbed SN power law.  The spectra were binned for visual clarity to five counts per bin.}}
\caption{\footnotesize{Fits for SN~2018bsz. The top panels show the spectra (points) and fitted model (thicker lines), the middle panels show the full model (and different model components for XMM), and the bottom panels show the log ratio of data to model. \label{fig:spectra} }}
\end{figure*}

The PN, MOS1 and MOS2 datasets are fitted simultaneously, with both $\Gamma$ and normalizations tied across instruments but allowing for 8\% relative variation \citep{2014Read} through a multiplicative constant (fixed to 1 for PN, free between 0.92--1.08 for MOS1 and MOS2). Table~\ref{tab:results_xmm} includes the resulting luminosities and limits for the power-law model in the upper panel, while Figure~\ref{fig:spectra} (b) shows the spectral fit corresponding to the top row of Table~\ref{tab:results_xmm}. For the best-fit level of contamination, the unabsorbed SN luminosity is constrained at $3\sigma$ in the 0.5--10~keV band (but not in 2--10~keV), implying a detection in the full band. The SN is also detected in the case of minimal contamination as expected, while we only obtain an upper limit on the SN luminosity in the maximal contamination scenario. The soft $\Gamma$ for all contamination scenarios suggest a significant softening of the SN spectrum since the early Chandra observations, with luminosities comparable or slightly lower at the time of the XMM observation.

Although CSM-ejecta interaction is often invoked for SLSNe-II and SLSNe-IIn in particular, it may also be relevant for SLSNe-I \citep{2017Yan,2018Lunnan,2022Pursiainen}. Motivated by the soft $\Gamma$ of the late XMM spectra, we additionally fit a \texttt{mekal} model alongside the contamination component, giving a total model of \texttt{tbabs(pow$_{\rm Cont.A}$ + mekal)}. Unlike the Chandra spectra, the late-time XMM spectra allow us to constrain the plasma temperature when adopting the best-fit $N_{\rm Cont.A}$ (Table~\ref{tab:results_xmm}, first row of the lower panel), yielding in a soft spectrum with $kT\sim0.2~$keV. We note, however, that only in the minimal contamination scenario can we constrain the 0.5--10~keV luminosity, but with a large upper bound for $kT$, which is outside the spectral range and therefore equivalent to unconstrained.

\begin{deluxetable*}{l l l l l l l}
\tablenum{3}
\tablecaption{Spectral Results for XMM EPIC Data \label{tab:results_xmm}}
\tablewidth{0pt}
\tablehead{ 
 \colhead{$N_{\Gamma \rm,cont.A}$}   & \colhead{$\Gamma$} & \colhead{C-stat}& \colhead{dof}& \colhead{$L_{\rm0.5-10~keV}$} & \colhead{$L_{\rm2-10~keV}$}\\
 \colhead{($\times10^{-7}$)}  &  \colhead{}& \colhead{}& \colhead{}& \colhead{($10^{40}~\rm erg~s^{-1}$)} & \colhead{($10^{40}~\rm erg~s^{-1}$)}
}
\startdata
        Average Cont.A 8.768  & $4.0_{-1.9}$ & 105.89 & 82 &$0.85_{-0.49}^{+2.12}$ & $<0.62$\\
        Min Cont.A 4.796    & $2.7_{-1.0}^{+1.4}$ & 105.71 & 82 &$1.10_{-0.44}^{+0.58}$ & $0.33_{-0.27}^{+0.63}$\\
        Max Cont.A 14.50    & $8.9_{-4.7}$ & 107.16 & 82 &$<4.91$ & $<3.57\times10^{-4}$\\[2mm]
        \hline
        & \colhead{$kT$} & & & & \\
        & \colhead{(keV)} & & & & \\
        \hline
        Average Cont.A 8.768  & $ 0.20_{-0.11}^{+0.24}$ & 107.32 & 82 &$<5.24$ & $<1.57$\\
        Min Cont.A 4.796    & $3.5_{-1.8}^{+43.7}$ & 108.38 & 82 &$1.02_{-0.50}^{+0.85}$ & $<2.16$\\
        Max Cont.A 14.50    & $0.17^{+0.20}$ & 107.89 & 82 &$<5.33$ & $<1.40\times10^{-4}$\\[2mm]
        \hline
\enddata 
\tablecomments{\footnotesize{The spectra were fitted with \texttt{const$\times$tbabs(pow$_{\rm Cont.A}$ + pow)} in the upper panel and \texttt{const$\times$tbabs(pow$_{\rm Cont.A}$ + mekal)} in the lower panel. For both models, \texttt{pow$_{\rm Cont.A}$} is the power law of the contamination annulus with frozen $\Gamma_{\rm Cont.A}$(=1.07) and normalization ($N_{\Gamma \rm,Cont.A}$). In the top row of both panels, we use the best-fit(average) value of $N_{\Gamma \rm,Cont.A}$, in the middle row, the lower 90\% bound (min) of $N_{\Gamma \rm,Cont.A}$ and in the bottom row the 90\% upper bound(max). The SN luminosities presented in the upper panel are derived from \texttt{pow} and over \texttt{mekal} in the lower panel and are given with 90\% error bars or as $3\sigma$ upper limits. All luminosities have been corrected for Galactic absorption. A distance of 111~Mpc is adopted and the epoch is 1253~days. }
}
\end{deluxetable*}

\section{Discussion}\label{sec:discussion}
We detect SN~2018bsz in five observations spanning 87 to 1253~days after explosion, with the final XMM observation requiring additional caution in its interpretation due to the nearby contaminating sources. This makes SN~2018bsz the only other X-ray detected SLSN-I besides the unusually bright SLSN-I SCP06F6 \citep{2013Levan}. The SN luminosity remains nearly constant as seen in Figure~\ref{fig:pulsar_evolution}, while the spectrum evolves from the very hard power law seen in the early Chandra data to a softer spectrum at later times. Although the $\Gamma$ of the late XMM spectrum is weakly constrained, the low plasma temperature obtained from alternative fits likewise supports a softer spectral shape (Table~\ref{tab:results_xmm}). 

In the following sections, we explore the two commonly invoked powering mechanisms for SLSNe: spin-down of a rapidly rotating magnetar and interaction between the ejecta and CSM, and assess their ability to account for the X-ray emission in SN~2018bsz.

\subsection{Magnetar origin} \label{sec:pulsar_discussion}
Millisecond magnetars are often invoked as the power source of SLSNe-I  (e.g. \citealt{2010Kasen,2010Woosley}). We therefore test whether the X-ray results of SN~2018bsz are compatible with such an engine.  

We use the common pulsar spin-down model, where pulsars are modeled as rotating dipoles in vacuum \citep{1969Ostriker,1983Shapiro}. The spin-down energy is given by
\begin{equation}\label{eqn:Erot_dot}
    \dot{E}_{\rm rot}=-4\pi^2I\frac{\dot{P}}{P^3} \hspace{5mm},
\end{equation}
where $I ~(=10^{45}~\rm g~cm^{2})$ is the moment of inertia, $P$ is the spin period, and $\dot{P}$ its time derivative. $P$ evolves as

\begin{equation}\label{eqn:p}
    P=\left[ P_0^2+\left(\frac{16\pi^2R^6B^2}{3Ic^3}\right)t\right]^{1/2},
\end{equation}
where $B$ is the magnetic field, which is assumed constant with time; $P_0$ is the pulsar spin down at birth; $c$ is the speed of light; and $R ~(=12~\rm km)$ is the radius of the neutron star. We adopt the magnetic field prescription from \cite{2016Shibata},

\begin{equation}\label{eqn:b_field}
    B^2= \frac{3}{2}\frac{Ic^3}{(2\pi)^2R^6}P\dot{P},
\end{equation}

which allows us to express $\dot{E}_{\rm rot}$ directly as a function of $P_0$ and $B$. A fraction $\eta=L_{\rm X}/\dot{E}_{\rm rot}$ of the spin-down power is assumed to emerge as X-ray emission in the 2--10~keV range. Since the X-ray efficiency of young pulsars and magnetars is poorly constrained, we adopt the empirically derived Crab value, $\eta=0.022$. This value was obtained from $\dot{E}_{\rm rot}=4.6\times10^{38}~\rm erg~s^{-1}$ \citep{2016Ansoldi} and the X-ray luminosity, which was derived from an extrapolated power law with $\Gamma=2.106$ \citep{2017Madsen}, extrapolated from the energy range from 3--50~keV into 2--10~keV.

To constrain $P_0$ and $B$, we run a Markov Chain Monte Carlo (MCMC) fitter from the \texttt{emcee} package \citep{2013Foreman}, using the pulsar model above. The fit accounts for asymmetric luminosity uncertainties by using a modified log-likelihood function, which converts the reported 90\% confidence intervals and 3$\sigma$ upper limit to equivalent 1$\sigma$ values. We run the fit over the 2--10~keV Chandra luminosities and XMM limit (black filled markers in Figure~\ref{fig:pulsar_evolution}), where the XMM limit corresponds to the average Cont.A results. We assume no ejecta absorption in these fits, which could be due to either ejecta asymmetries or ionization of the ejecta from the magnetar, discussed further below. Furthermore, we assume flat initial priors in log space ($\log_{10}(P_0)$ and $\log_{10}(B)$), with $\rm0.5~ms< P_0<1~s$ and $\rm 10^{11}~G<B<10^{16}~G$. The MCMC fit is initialized with 1000 walkers, evolved over 10,000 steps, with the first 200 discarded as burn-in.

\begin{figure*}
\hspace{0cm}\includegraphics[width=1\textwidth]{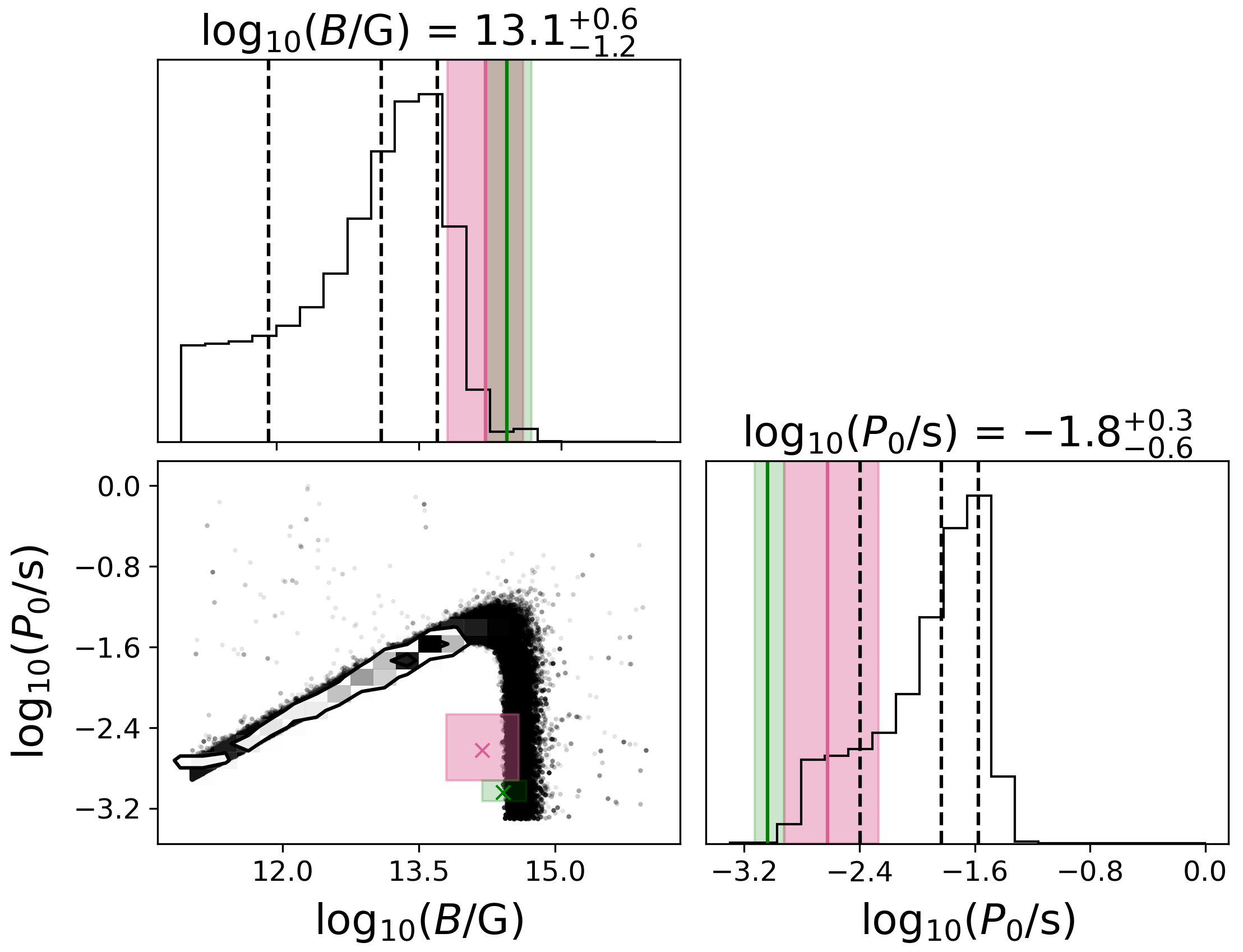} 
\caption{ Corner plot of the MCMC run for $B$ and $P_0$. The black dashed lines in the top left and bottom right show the median and 1$\sigma$ ranges for the best fit parameters, which are given at the top of respective histogram. The contour lines in the bottom left panel represent 1$\sigma$ and 2$\sigma$. The green regions correspond to $BP_{\rm optical}$ with 1$\sigma$ distribution and best-fit optical values marked as a dark-green cross and lines.  The pink lines and shaded regions correspond to the median $B$ and $P_0$ with corresponding 1$\sigma$ distribution from the results of the full sample of 262 SLSNe by \cite{2024Gomez}.
\label{fig:corner_plot}}
\end{figure*}

The MCMC fit results in $\log B=13.1^{+0.6}_{-1.2}$ and $\log P_0=-1.8^{+0.3}_{-0.6}$ (corresponding to $P_0\sim16~\rm ms$), as shown in the corner plot in Figure~\ref{fig:corner_plot}.  The fit is also illustrated by the solid black line in Figure~\ref{fig:pulsar_evolution}. If we instead use the XMM limit from min Cont.A, we get a similar result with $\log B\sim13.5$ and $\log P_0\sim-1.6$ and a good fit to the data. Using the max Cont.A value yields a fit that deviates substantially from the observed data, resulting in parameter estimates of $\log B\sim 14.9$ and $\log P_0\sim-1.0$. 

We compare these results to those inferred from fits of the optical light curve of SN~2018bsz by \cite{2024Gomez}, who found best-fit values of $P_0=0.9_{-0.2}^{+0.3}$~ms and $\log B=14.4_{-0.2}^{+0.3}$ (from here on, we refer to these best-fit values values as $BP_{\rm optical}$), with corresponding $1\sigma$ confidence interval. We use $BP_{\rm optical}$ in our pulsar model and include the results in Figure~\ref{fig:pulsar_evolution} and Figure~\ref{fig:corner_plot} as a green line for the best-fit values and green shaded regions for the 1$\sigma$ confidence interval. From the corner plot in Figure~\ref{fig:corner_plot}, we see that the MCMC results are generally not in line with $BP_{\rm optical}$ (green region) no with the sample results from light curve fits of 262 SLSNe from \cite{2024Gomez} (pink region), including SN~2018bsz. The corner plot also reveals a strong degeneracy between $P_0$ and $B$ (bottom left panel of Figure~\ref{fig:corner_plot}). Although the $B-P_0$ degeneracy allows for some overlap with $BP_{\rm optical}$ at larger $B$, these solutions lie outside the 2$\sigma$ contours and ar disfavored (see bottom left panel of Figure~\ref{fig:corner_plot}). A larger $B$, such as suggested by $BP_{\rm optical}$, is clearly not compatible with the temporal evolution of our observational results, as seen in Figure~\ref{fig:pulsar_evolution}. The observed light curve requires a lower $B$ for a flatter evolution.

We also explore whether a lower $\eta$ could reconcile the X-ray light curve with $BP_{\rm optical}$ by repeating the MCMC with fixed $B$ and $P_0$ at $BP_{\rm optical}$ and fitting for $\eta$ with flat priors ($10^{-6}<\eta<1$), we obtain $\eta=0.01$. This value is below that of the Crab and smaller than typical efficiencies for pulsars in X-rays (e.g., \citealt{2004Cheng,2008Li}), but foremost, it does not change the temporal luminosity evolution, which is the main discrepancy between our MCMC results and $BP_{\rm optical}$.

\subsubsection{Impact of ejecta absorption}
The X-ray luminosities derived from our spectral fits assume negligible ejecta absorption, since the low spectral count rates prevent us from fitting for intrinsic absorption directly and the spectra do not show strong absorption features (see Figure~\ref{fig:spectra}). Such an assumption of no ejecta absorption is not necessarily expected a priori. The ejecta masses inferred for SLSNe-I are typically in the range of $2-30~M_\Sun$ \citep{2015Nicholl,2017Nicholl,2017Jerkstrand}, where such ejecta are generally expected to remain optically thick to soft X-rays for a significant time after explosion. For SN~2018bsz specifically, \cite{2024Gomez} inferred an ejecta mass of $M_{\rm ej}=9.5_{-4.4}^{+7.5}~ M_\Sun$ based on optical light-curve modeling.

The apparent lack of strong absorption signatures may indicate that the magnetar is in fact fully absorbed and that the X-ray emission is due to CSM interaction (a scenario discussed in Section~\ref{sec:CSM}). This could also explain the discrepancy between the $B,P_0$ inferred from optical data and our X-ray data. Alternatively, the absorption of the magnetar emission is very low, either due to geometric asymmetries of the ejecta or because the ejecta have become highly ionized by radiation from the magnetar. In the latter case, the opacity to soft X-rays can be reduced as the PWN progressively ionizes the surrounding ejecta \citep{2014Metzger}. Once the ionization front reaches the outer ejecta surface, X-rays can escape freely, a process referred to as ionization breakout. A related possibility is that the emission comes from an extended PWN, in which case the observed spectrum may represent a superposition of emission components with different $\Gamma$ and absorption, which could further obscure clear absorption signatures. 

Following \cite{2018Margutti}, we estimate the ionization breakout time ($t_{\rm ion}$) and the corresponding X-ray luminosity at $t_{\rm ion}$ using the magnetar parameters $BP_{\rm optical}$ together with the ejecta mass $M_{\rm ej}=9.5_{-4.4}^{+7.5}~M_\Sun$ and velocity $v_{\rm ej}=15,127^{+1873}_{-2345}\rm km~s^{-1}$ inferred for SN~2018bsz by \cite{2024Gomez}. For this parameter set, the ionization breakout occurs at $t_{\rm ion}\approx 72~\rm yr$ with $L_{\rm x}(t_{\rm ion})\approx1.6\times10^{37}~\rm erg~s^{-1}$. Adopting the most favorable combination within the quoted $1\sigma$ confidence intervals of $BP_{\rm optical}$, $M_{\rm ej}$ and $v_{\rm ej}$, corresponding to the lowest $B$ and lowest ejecta absorption, yields $t_{\rm ion}\approx 5~\rm yr$ with $L_{\rm x}(t_{\rm ion}) \approx 9.3\times10^{39} ~\rm erg~s^{-1}$. In both cases, the predicted ionization breakout occurs beyond the temporal window of our observations, though with a detectable flux of $6.3\times10^{-15}~\rm erg~s^{-1}~cm^{-2}$ for the most favorable scenario, at a distance of 111~Mpc. Given the late $t_{\rm ion}$, it is unlikely that the X-ray emission is due to ionization breakout from a magnetar with $BP_{\rm optical}$ properties. 

We note, however, that the ionization breakout models are one-dimensional and assume spherical symmetry. In more realistic scenarios, asymmetric ejecta structures may lead to low-density channels with reduced absorption, giving an asymmetric ionization breakout \citep{2025Eiden}. Such effects could, in principle, relax the constraints derived from spherically symmetric models and partially reconcile a magnetar-powered scenario with the observed luminosities. However, reproducing both the high X-ray luminosities and the relatively flat temporal evolution would still require a finely tuned combination of magnetar parameters and ejecta properties. Even when accounting for asymmetric ionization and reduced ejecta absorption, it remains challenging for a magnetar with $BP_{\rm optical}$-like properties to account for the observed X-ray evolution.

\begin{figure}
\hspace{-0.2cm}\includegraphics[width=0.49\textwidth]{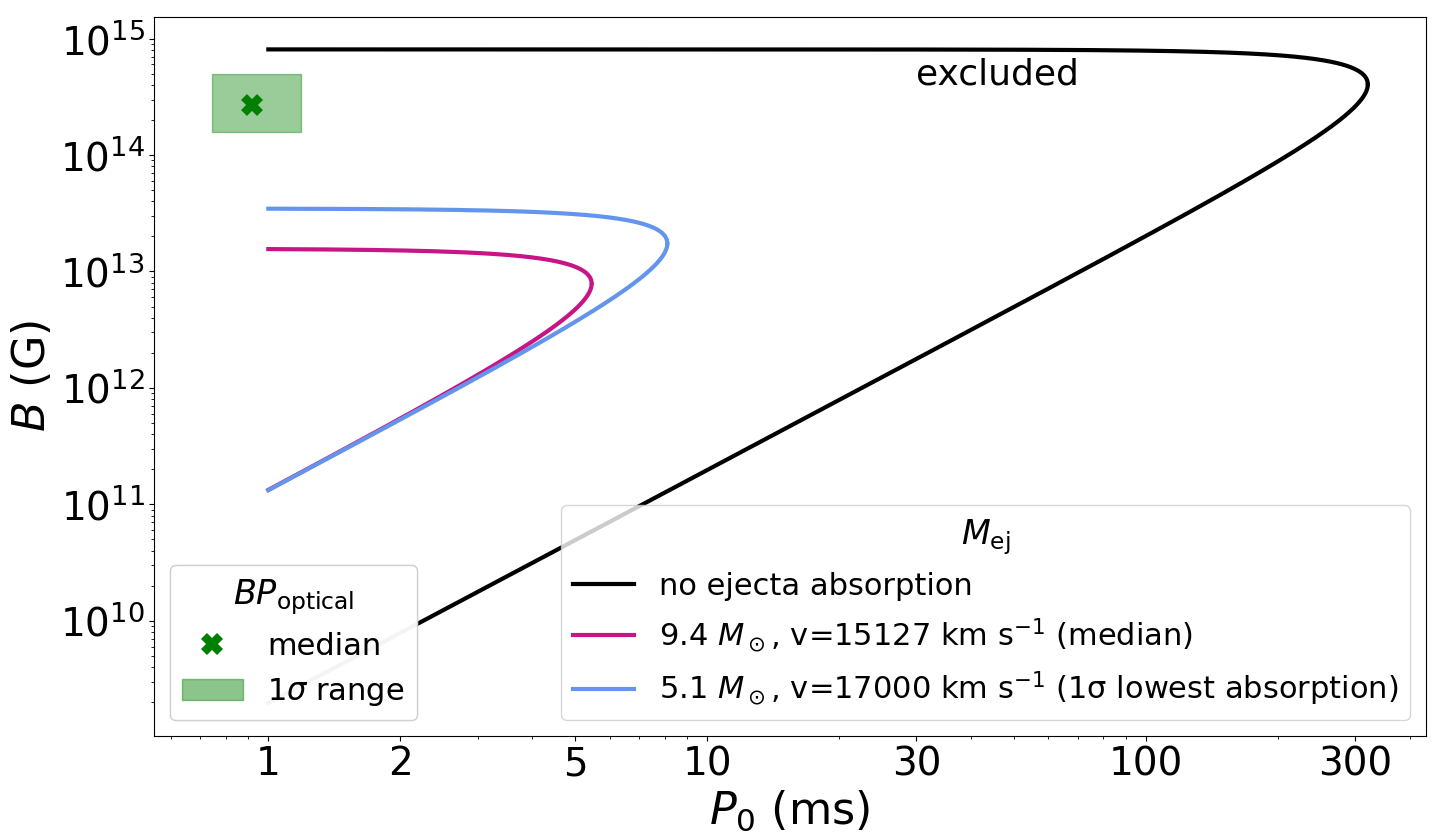} 
\caption{Excluded pulsar/magnetar parameter space for three cases: negligible ejecta absorption (black), ejecta properties corresponding to the best-fit $M_{\rm ej}$ and $v_{\rm ej}$ inferred from optical data for SN~2018bsz (magenta; \citealt{2024Gomez}), and a minimum-absorption scenario defined by the lowest $M_{\rm ej}$ and highest $v_{\rm ej}$ within the $1\sigma$ confidence intervals (blue; \citealt{2024Gomez}). All lines are derived from the XMM luminosity limit in 2--10~keV assuming the average Cont.A. The black line is derived from the pulsar models in Equations \ref{eqn:Erot_dot}--\ref{eqn:b_field}, and the others are based on ionization breakout following Equations~1 and 4 of \cite{2018Margutti}. The green cross shows the $BP_{\rm optical}$, and the green box the corresponding 1$\sigma$ spread. 
\label{fig:BP_plot}}
\end{figure}

Figure~\ref{fig:BP_plot} illustrates the excluded parameter space in the $B-P_0$ plane based on the 2--10~keV luminosity limit from the XMM observation (best-fit Cont.A), which provides the strongest constraints due to the late epoch. The figure shows excluded regions for the case of negligible ejecta absorption and for two examples of absorption by ejecta with  $M_{\rm ej}$ and $v_{\rm ej}$ within the constraints from \cite{2024Gomez}. The curve corresponding to no ejecta absorption (black) is derived using Equations~(\ref{eqn:Erot_dot})--(\ref{eqn:b_field}), while the curves including ejecta absorption are derived using the ionization-breakout formalism (Equations~1 and 4 of \citealt{2018Margutti}). We note that our adopted pulsar models differ somewhat from the setup used by \cite{2018Margutti}, resulting in spin-down luminosities higher by a factor of $\sim 3$.\footnote{This is due to the definition used for the spin-down dipole model by the angle between spin axis and magnetic field, as well as the assumed $\eta$.}  This analysis excludes pulsars born with low $B$ and short $P_0$, which would have fully ionized the ejecta at earlier times and produced significantly higher luminosities than observed. Magnetar properties inferred for SN~2018bsz by \cite{2024Gomez} (green cross and shaded region in Figure~\ref{fig:BP_plot}) are consistent with the observations for expected levels of X-ray absorption.

We note, however, that the complex optical light curve implies that there are likely systematic uncertainties on $BP_{\rm optical}$ that are not fully captured by the shaded region in Figure~\ref{fig:BP_plot}. Using the online data provided by \cite{2024Gomez}, we reconstruct the light curves, since the figure included in the paper is too small for a detailed evaluation. We find that the post-peak plateau is not fully captured by the fit, which may indicate that additional physical processes beyond magnetar powering contribute to the observed light curve. This interpretation is consistent with \cite{2022Pursiainen}, who suggested that the post-peak plateau could originate from an optically thick, H-rich CSM.

In summary, the temporal evolution of the observed X-ray emission is not consistent with a magnetar central engine characterized by $BP_{\rm optical}$ but instead requires a lower $B$ to reproduce the flatter luminosity evolution. While the earliest Chandra observation could be marginally compatible with $BP_{\rm optical}$ in the presence of little absorption, the later observed luminosities exceed the predictions of the magnetar model. A more plausible explanation is that the X-ray emission is dominated by interaction between the ejecta and CSM.

\subsection{CSM interaction}\label{sec:CSM}
An alternative energy source proposed to explain the brightness of some SLSNe-I (e.g. \citealt{2012Leloudas,2013Chatzopoulos,2017Yan}), including SN~2018bsz \citep{2022Pursiainen}, is the ejecta-CSM interaction. For SN~2018bsz, \cite{2022Pursiainen} identified hydrogen emission lines emerging $\sim10-40~\rm days$ after peak optical luminosity, with spectral evolution more reminiscent of common SNe IIn than other SLSNe-I, interpreted as interaction with an asymmetric, disk-like CSM. There are also some SLSNe-I which show similar behaviors, where the spectra develop H emission at $\gtrsim100~$days, indicating presence of CSM interaction \citep{2015Yan,2017Yan,2023Margutti}. SN~2017ens is such an example, which showed H-line emission and CSM interaction that was detected in radio at $\sim1250~$days \citep{2018Chen,2023Margutti}. These signatures suggested a dense shell-like and H-rich CSM structure.

Our results for SN~2018bsz could likewise be interpreted as signatures of strong early CSM interaction. The early Chandra spectra are best described by a hard power law with $\Gamma=1.32$ (see Table~\ref{tab:results_chandra} for details), whereas by the XMM epoch, the spectrum has softened to $\Gamma<4.0$ and is equally well fit by a thermal plasma model with $kT=0.20~\rm keV$ (Table~\ref{tab:results_xmm}). The spectral softening is significant at $2\sigma$ (assuming average Cont.A). This behavior is consistent with an initially strong interaction that weakens with time (e.g., \citealt{2016Dwarkadas,2017Chevalier}).
The little to no decline in luminosity further favors a CSM interaction scenario over a magnetar powered one, which would be expected to fade more rapidly. We note, however, that the inferred XMM luminosity depends on the assumed level of contamination; in the maximal-contamination case, a significant decrease is observed. Although our $\Gamma$ uncertainties are large or in some cases only lower limits (Table~\ref{tab:results_xmm}), we also find constrained plasma temperatures indicative of a soft spectrum at late times. Similar spectral softening in X-rays and late-time hydrogen-line emergence have been observed in ordinary stripped type SNe transitioning into CSM-interacting systems (e.g. \citealt{2006Soderberg,2015Milisavljevic,2018Chen,2018Mauerhan,2022Brethauer}).

The inferred X-ray luminosities and limits of SN~2018bsz are consistent with the highest luminosities expected from CSM interaction as suggested by \cite{2013Levan}, assuming a high mass-loss rate of $\sim10^{-4}~M_\Sun~\rm yr^{-1}$. Higher X-ray luminosities would require larger ejecta masses and higher mass-loss rates. Using the mass-loss-luminosity relation from \cite{2017Chevalier}, we estimate the progenitor mass-loss rate based on each observation, using the formula:

 \begin{equation}
    Lx\approx 3\times 10^{39}g_{ff}C_n\left(\frac{\dot{M}_{-5}}{v_{w1}}\right)^2t_{10}^{-1},
\end{equation}

where $C_n=(n-3)(n-4)^2/[4(n-2)]$ for the reverse shock (adopting $\rm n=7$), $g_{ff}$ is the Gaunt factor in order unity, and $\dot{M_{5}}$ is the mass-loss rate in units of $10^{-5}~M_\Sun~\rm yr^{-1}$. The wind velocity, $v_{w1}$, is in units of 10~km~s$^{-1}$, which we adopt for stripped type SNe progenitors with typical value of $\sim 1000~\rm km~s^{-1}$ \citep{2014Smith,2017Chevalier}. Finally, $t_{10}$ is time in units of 10~days. We stress that the estimated $\dot{M_{5}}$ are only indicative as these calculations assume a steady wind and rely on additional assumptions, such as $v_{w1}$ and $n$. 

The resulting mass-loss rates for the Chandra epochs averages to $\sim 5.4\times10^{-3}~M_\Sun~\rm yr^{-1}$ and for the XMM epoch $\sim 1.4\times10^{-2}~M_\Sun~\rm yr^{-1}$ adopting the luminosity from the thermal fit with the average Cont.A. These values exceed those of typical ordinary stripped SNe \citep{2014Smith,2017Chevalier}, most SLSNe-I \citep{2018Margutti} and that of SN~2017ens \citep{2018Chen,2023Margutti}, and are closer to some mass-loss rates for SLSNe-II \citep{2007Smith}, though the ranges for SLSNe-II often reach $\gtrsim 0.01~M_\Sun~\rm yr^{-1}$ \citep{2019Chatzopoulos,2025Pessi}. Our mass-loss rates are most comparable to those of the subset of stripped SNe that evolve into strong interacting systems \citep{2022Brethauer}. It is noteworthy that multiwavelength observations of SN~2018bsz already indicate atypical behavior for an SLSN-I, including an early optical plateau and hydrogen emission \citep{2018Anderson,2022Pursiainen}. This agrees well with the slightly larger mass-loss rates found compared to other SLSNe-I, and could have been ejected as previously discussed, in a violent eruption of the progenitor star.

Given the derived mass-loss rates, we estimate the time of the progenitor mass loss to assess whether the X-ray detections from Chandra could originate from the disk-like CSM structure proposed in \cite{2022Pursiainen}.  We use $t_{\rm CSM}=t_{\rm Xray,obs}v_{\rm shock}/v_w$ and estimate the shock velocity ($v_{\rm shock}$) for a spherically symmetric wind following \cite{2017Chevalier}. We adopt the same assumptions for $n$ and $v_w$ as previously, as well as the values of $E_{\rm ej}$ and $M_{\rm ej}$ for SN~2018bsz from \cite{2024Gomez}. This yields $t_{\rm CSM}\approx 5-18$~yr before explosion, implying a mass-loss episode lasting for $\sim13~\rm yr$. The corresponding CSM radius is $R_{\rm CSM\approx (1.7-5.8)\times10^{16}~\rm cm}$. For comparison, \cite{2022Pursiainen} derived an upper limit for the disk-like CSM radius of 
$<6.5\times10^{15}~\rm cm$. While our estimated radius is larger, there are many simplifying assumptions, including considering spherical symmetry rather than a disk-like geometry, and we conclude that the X-ray emission is compatible with an origin in the CSM disk proposed by \cite{2022Pursiainen}.

\section{Summary and Conclusion}\label{sec:summary_conclusion}

We detect X-ray emission from SN~2018bsz in Chandra observations spanning 87 to 304~days after explosion at $3\sigma$ significance, with an average luminosity of $L=1.1\times 10^{40}~\rm erg~s^{-1}$ in the 0.5--8~keV energy range. A possible detection is also seen in the 0.5--10~keV XMM observation at 1253~days post explosion; however, this result is sensitive to assumptions about nearby contaminating sources and should be interpreted with caution. Independent of the XMM detection, SN~2018bsz is the second SLSN-I detected in X-rays, and the third of all SLSNe detected in X-rays. It exhibits the lowest X-ray luminosity among this small sample.

Interpreted within a magnetar-powered framework and assuming negligible ejecta absorption, the observed X-ray luminosities favor magnetar parameters with a comparatively weak magnetic field ($\log B\sim$ 13.1) and a long initial spin period ($P_0\sim16~\rm ms$). These values lie outside the range typically inferred for millisecond magnetars (see, e.g., \citealt{2017Nicholl,2024Gomez}), and are difficult to reconcile with the energetics required to power the superluminous optical emission.

Adopting instead the magnetar parameters inferred from the optical light curve by \cite{2024Gomez}, we find that the predicted X-ray luminosities are lower than the observed values at all but the earliest epoch and fail to reproduce the relatively flat temporal evolution. Accounting for ejecta absorption further increases this discrepancy. While asymmetric ejecta and ionization effects may reduce the effective absorption along the line of sight, our estimated timescale of ionization breakout occurs at best $\sim2$~yr after our observational window assuming $BP_{\rm optical}$, though at an observable flux ($6.3\times10^{-15}~\rm erg~s^{-1}~cm^{-2}$). Although such effects could partially alleviate the tension, reproducing both the X-ray luminosity level and temporal evolution would be difficult given the millisecond magnetar properties of $BP_{\rm optical}$.

Taken together, these results disfavor a standard magnetar-powered origin for the observed X-ray emission. Instead, the data are more readily explained by strong early ejecta-CSM interaction. The X-ray spectrum is initially hard at epochs 87--304~days, with a photon index of $\Gamma=1.32$, and softens substantially by the XMM observation at 1253~days, where it is consistent with either $\Gamma<4.0$ or a thermal plasma temperature of $kT=0.20~\rm keV$. This spectral evolution, combined with the relatively slow luminosity decline, is consistent with a scenario in which strong early interaction weakens over time.

From the observed X-ray luminosities, we estimate progenitor mass-loss rates of order a few $10^{-3}~ M_\Sun \rm ~yr^{-1}$. These values exceed those typically inferred for ordinary stripped SNe and SLSNe-I and are more comparable to the subset of stripped SNe which evolve into strongly interacting systems following eruptive mass-loss episodes that produce dense, shell-like CSM.

The presence of CSM interaction and elevated mass-loss rates is consistent with previous multi-wavelength studies of SN~2018bsz \citep{2018Anderson,2021Chen,2022Pursiainen}, which revealed hydrogen emission lines and evidence for interaction with a disk-like CSM \citep{2022Pursiainen}. Together, these results suggest that SN~2018bsz represents a distinct subclass within the SLSN-I population, where CSM interaction plays a significant role in shaping the high-energy emission, and where a magnetar central engine, if present, is unlikely to dominate the observed X-ray output.

\begin{acknowledgments}
We thank the anonymous referee for the comments, which helped improve the manuscript. This work was supported by the Knut and Alice Wallenberg Foundation. This research has made use of data from the Chandra Data Archive and Chandra Source Catalog, both provided by the Chandra X-ray Center (CXC) and employs a list of Chandra datasets, obtained by the Chandra X-ray Observatory, contained in the Chandra Data Collection (CDC) ~\dataset[doi:10.25574/cdc.476]{https://doi.org/10.25574/cdc.476}.This research has made use of software provided by the Chandra X-ray Center (CXC) in the application package CIAO. Based on observations obtained with XMM-Newton, an ESA science mission with instruments and contributions directly funded by ESA Member States and NASA. This research has made use of software provided by the High Energy Astrophysics Science Archive Research Center (HEASARC), which is a service of the Astrophysics Science Division at NASA/GSFC. 

Funded by the European Union (ERC, project number 101042299, TransPIre). Views and opinions expressed are, however, those of the author(s) only and do not necessarily reflect those of the European Union or the European Research Council Executive Agency. Neither the European Union nor the granting authority can be held responsible for them.

\facilities{CXO(ACIS), XMM (EPIC-PN)}

\software{CIAO/CALDB \citep{2006SCIAO},
          HEAsoft \citep{2014Nasa},
          SAOImage DS9 \citep{2003SAOImage},
          \texttt{astropy} \citep{2013astropy,2018astropy,2022astropy},
          \texttt{numpy} \citep{2020numpy},
          \texttt{scipy} \citep{2020SciPy},
          XSPEC \citep{1996Arnaud}, \texttt{corner.py} \citep{corner}
          }
\end{acknowledgments}

\bibliography{ref}{}
\bibliographystyle{aasjournal}

\end{document}